\documentclass[11pt,twoside]{article}
\usepackage{maxim}

\begin{document}
\pagestyle{myheadings}
\markboth{Maxim Raginsky}{Quantum System Identification}
\begin{titlepage}
\title{{\bf Quantum System Identification}\footnote{Paper for the
    invited session ``Quantum measurement, filtering, and quantum
    feedback control'' (V.P. Belavkin, organizer) at the {\it International Conference ``Physics and
    Control''} ({\it PhysCon 2003}), August 20-22, St.~Petersburg, Russia.}}
\author{Maxim Raginsky\thanks{Electronic Mail:
    \texttt{maxim@ece.northwestern.edu}
}
  \\[1ex]
  {\small Center for Photonic Communication and Computing}\\
  {\small Department of Electrical and Computer Engineering}\\
  {\small Northwestern University, Evanston, IL 60208-3118, USA}}
\date{}

\maketitle

\begin{abstract}
\noindent{We formulate and study, in general terms, the problem of quantum
  system identification, i.e., the determination (or estimation) of
  unknown quantum channels through their action on suitably chosen
  input density operators. We also present a quantitative analysis of
  the worst-case performance of these schemes.}
\end{abstract}

\setcounter{page}{0}
\thispagestyle{empty}
\end{titlepage}

\section{Introduction and background}\label{sec:intro}

In quantum information theory \cite{key} all admissible devices are
described mathematically by means of the so-called {\em quantum
  operations} (or {\em quantum channels})
\cite{dav,kra}. 

Given a complex Hilbert space $\cH$, denote by $\cB(\cH)$ the
$*$-algebra of all bounded operators on $\cH$. In this paper we will
work primarily with finite-dimensional Hilbert spaces, so that
$\cB(\cH)$ includes all linear operators on $\cH$. Given Hilbert
spaces $\cH_1$ and $\cH_2$, a quantum channel $T$ is a completely
positive trace-preserving linear map of $\cB(\cH_1)$ into
$\cB(\cH_2)$. All such maps admit the {\em Kraus decomposition} \cite{kra}
\begin{equation}
T(\rho) = \sum_k A_k \rho A^*_k,
\label{eq:krausform}
\end{equation}
where $\map{A_k}{\cH_1}{\cH_2}$ are operators satisfying $\sum_k A^*_k
A_k = \idty_{\cH_1}$. This definition of the quantum channel is
formulated in the \Schrodinger\ picture, so that the density operators
on $\cH_1$ are mapped to density operators on $\cH_2$. The corresponding
Heisenberg-picture definition goes the other way (observables on
$\cH_2$ are mapped to observables on $\cH_1$) and yields a
completely positive unit-preserving linear map
$\map{\widehat{T}}{\cB(\cH_2)}{\cB(\cH_1)}$ related to $T$ by the
duality
\begin{equation}
\tr [T(\rho) X] = \tr [\rho \widehat{T}(X)]
\label{eq:duality}
\end{equation}
for all $\rho \in \cB(\cH_1)$ and all $X \in \cB(\cH_2)$. In this
paper we will deal mostly with the \Schrodinger\ picture.

This seemingly simple framework turns out to be rich enough to cover
all kinds of general transformations of quantum-mechanical states. In
fact, both purely classical and hybrid (classical-quantum or
quantum-classical) transformations can be included as well, simply by
restricting to a suitable Abelian subalgebra either at the input or at
the output.

One of the basic challenges, both for theoreticians and for
experimentalists, is to discover efficient procedures for analysis and
synthesis of quantum channels. For instance, when designing a device
for a specific task (e.g., an optimal quantum cloner \cite{wer}), one has to
run tests in order to determine whether the device performs according
to specification. Several such procedures have been proposed already,
such as the tomographic scheme of D'Ariano and Lo Presti \cite{dlp} or
the maximum-likelihood reconstruction method of Je\v{z}ek, Fiur\'a\v{s}ek,
and Hradil \cite{jfh}.

All of these schemes rely, in one way or
another, one the one-to-one correspondence \cite{jam} between
completely positive maps $\cB(\cH_1) \rightarrow \cB(\cH_2)$ and
positive operators on $\cH_2 \tp \cH_1$, to which we shall return
later in this paper. Our purpose here is to phrase the ideas
common to these schemes as an abstract problem of {\em system
  identification}.

\section{The quantum system identification problem}\label{sec:qsysid}

Consider the following arrangement, shown in Fig.~\ref{fig:io}: we are
given a ``black box'' that implements an unknown quantum channel $\map{T}{\cB(\cH_1)}{\cB(\cH_2)}$,
which we need to determine. This will be done by presenting to the
black box certain suitably chosen input density operators $\rho$,
thereby obtaining output density operators $\sigma \equiv T(\rho)$,
and then trying to determine (or to estimate) $T$ given a set of
ordered pairs $\left(\rho,T(\rho)\right)$.

\begin{figure}
\begin{center}
\begin{picture}(220,80)   
\put(8,24){$\rho$}
\put(0,14){\small \sf Input}
\put(15,26){\vector(1,0){18}}
\put(44,38){\line(1,0){10}}
\put(35,23){$\left \{ \begin{array}{l}
\, \\
\,
\end{array} \right.$}
\put(44,14){\line(1,0){101}}   
\put(55,25){\framebox(80,30)
{\begin{minipage}[center]{0.6in}
\begin{center}
\small{
{\sf Quantum channel}
$T$}
\end{center}
\end{minipage}}}  
\put(135,38){\line(1,0){10}}
\put(135,23){$\left. \begin{array}{l}
\, \\
\,
\end{array} \right\}$}
\put(156,26){\vector(1,0){18}}
\put(177,24){$T \tp \id(\rho)$}
\put(182,14){\small \sf Output}
 \end{picture}
\end{center}
\caption{The set-up for the quantum system identification problem that
allows for the use of quantum entanglement.}\label{fig:subs}
\end{figure}
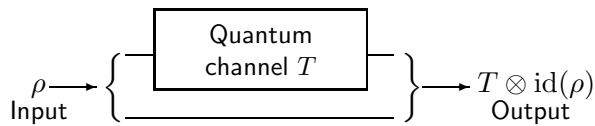

We assume that we can re-use the black box any finite number of times,
and that we can employ it as part of a more complicated
arrangement. A typical strategy \cite{dlp} is to use quantum
entanglement \cite{bo}: one prepares input states
(density operators) on the tensor product space $\cH_1 \tp \cH_1$, and
then subjects only one subsystem of the resulting composite system to
$T$ (see Fig.~\ref{fig:subs}). This way we have the channel $T \tp
\id$ from $\cB(\cH_1 \tp \cH_1)$ into $\cB(\cH_2 \tp \cH_1)$, which
corresponds uniquely to the original channel $T$. This extension is,
in fact, at the basis of the Jamio\l kowski isomorphism \cite{jam}
(see next Section for a detailed discussion).

The problem of {\em quantum system identification} can now be
formulated as follows. Instead of viewing the arrangement shown in Fig.~\ref{fig:subs} as a mapping of density operators $\rho$ to
density operators $T\tp \id(\rho)$, we can think of it as a mapping of
quantum channels $T$ into density operators $\rho[T]$ (we will use square brackets to distinguish the maps whose arguments are quantum
channels from the maps that take density operators as arguments). That
is, if we fix a density operator $\rho$, then we have $\rho[T] :=
T\tp \id(\rho)$. We will say that a density operator $\rho$ is {\em
  admissible} if the map $\rho[\bullet]$ is invertible. Given an
admissible density operator $\rho$, we will denote by $\rho^\sharp$ the
inverse mapping from density operators to channels.

This points, at least in principle, toward a solution of the problem
of quantum system identification. All we need to do is to prepare an
admissible state $\rho$, launch one of its subsystems through the
black box $T$ to get the output density operator $\sigma \equiv T\tp
\id(\rho)$, and then reconstruct $T$ as the inverse
$\rho^\sharp(\sigma)$. Of course now we are faced with (at least) two
more problems. (1) What states are admissible? (2) What can we say
about the performance of the reconstruction procedure as a function of
the (admissible) input state? We will address these problems in the
remainder of this paper. The rest of this section will be devoted to
discussion of the general properties of the map $\rho[\bullet]$.

\begin{figure}
\begin{center}
\begin{picture}(220,80)   
\put(10,28){\small \sf Input}
\put(28,38){\vector(1,0){36}}    
\put(20,38){$\rho$}
\put(64,25){\framebox(80,30)
{\begin{minipage}[center]{0.6in}
\begin{center}
\small{
\sf Quantum channel
$T$}
\end{center}
\end{minipage}}}  
\put(144,38){\vector(1,0){36}}      
\put(185,38){$T(\rho)$}
\put(180,28){\small \sf Output}
 \end{picture}
\end{center}
\caption{Input-output diagram for the quantum system identification
  problem.}\label{fig:io}
\end{figure}
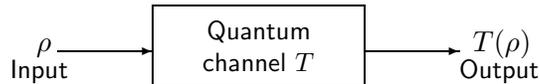

We would like to make some statements about the continuity of
$\rho[\bullet]$. Let us equip the algebra $\cB(\cH)$ of linear operators on the Hilbert
space $\cH$ with the {\em trace norm} \cite{rs}, defined by
$\trnorm{X} := \tr \abs{X}$, where the absolute value of an operator
$X$ is defined as $\abs{X}:=\sqrt{X^*X}$. Then $\trnorm{A} = \tr A$ for any
positive operator $A$, and furthermore $\trnorm{\rho}=1$ for any
density operator $\rho$. We will also need to estimate norm
differences of quantum channels; the ideal norm for this purpose is
the so-called {\em norm of complete boundedness} (or CB-norm, for
short) \cite{pau}, defined by
\begin{equation}
\cbnorm{T} := \sup_{n \in \N} \norm{T \tp \id_n}, \qquad
\map{\id_n}{\cM_n}{\cM_n}
\label{eq:cbnorm}
\end{equation}
where $\cM_n$ stands for the algebra of $n \times n$ complex
matrices. The norm $\norm{\bullet}$ on the r.h.s. of (\ref{eq:cbnorm}) is the
{\em operator norm}, defined for a general linear map
$\map{M}{\cB(\cH_1)}{\cB(\cH_2)}$ by
\begin{equation}
\norm{M} := \sup_{X \in \cB(\cH_1); \trnorm{X} \le 1}\trnorm{M(X)}.
\label{eq:opnorm}
\end{equation}
(Note that the above definition is tailored specifically for quantum
channels in the \Schrodinger\ picture; consult the monograph of
Paulsen \cite{pau} for generalities.) We have $\cbnorm{T} =1$ for any
quantum channel $T$. We shall have an occasion to use
some other properties of the CB-norm in later sections; all
we need right now is the inequality $\trnorm{T(A)} \le \cbnorm{T}\trnorm{A}$,
which is obvious from definitions, and the multiplicativity of the
CB-norm with respect to the tensor product, $\cbnorm{S \tp T} =
\cbnorm{S}\cbnorm{T}$.

With these preliminaries out of the way, consider a density operator
$\rho$. We can easily see that the map $\rho[\bullet]$ is
continuous. Indeed, consider two quantum channels, $T$ and $T'$. By
definition we have
\begin{eqnarray}
\trnorm{\rho[T] - \rho[T']} &\equiv & \trnorm{T\tp \id(\rho) - T'\tp \id(\rho)}
\nonumber \\
&\le & \cbnorm{T\tp\id-T'\tp\id} \nonumber \\
&=& \cbnorm{T - T'}
\label{eq:cont}
\end{eqnarray}
(we have used the fact that the trace norm of a density operator is
equal to one).

\section{Admissible states and the Jamio\l kowski isomorphism}\label{sec:jam}

In this section we describe an approach to the construction of
admissible states. This method is closely connected to the Jamio\l
kowski isomorphism \cite{jam}, and is, in fact, a natural
extension of the latter. First we need some mathematical machinery
from the theory of completely positive maps.

Given a C*-algebra $\cA$,
denote by $\cA^+$ the cone of positive elements of $\cA$ (i.e.,
precisely those elements that can be written in the form $A^*A$ for
some $A \in \cA$). A linear map $T$ between C*-algebras $\cA$ and
$\cB$ is called {\em positive} if $T(\cA^+) \subseteq \cB^+$, and
{\em completely positive} (CP, for short) if the maps $\map{T \tp \id_n}{\cA \tp \cM_n}{\cB
  \tp \cM_n}$ are positive for all $n \in \N$. A quantum channel is
thus a specific instance of a CP map.

According to the fundamental theorem of
Stinespring \cite{sti}, for any CP map
$\map{T}{\cB(\cH_1)}{\cB(\cH_2)}$ there exist a Hilbert space $\cE$
and a bounded operator $\map{V}{\cH_2}{\cH_1 \tp \cE}$, such that
\begin{equation}
T(\rho) = V^*(\rho \tp \idty_\cE)V.
\label{eq:sti}
\end{equation}
The pair $(V,\cE)$ is called the {\em Stinespring dilation} of
  $T$. Furthermore, with the additional requirement that the linear
  span of the set $\setcond{(A
  \tp \idty_\cE)V\psi}{A \in \cB(\cH_1),\psi \in \cH_2}$ be dense in
  $\cH_1 \tp \cE$, the pair $(V,\cE)$ determines the CP map $T$
  uniquely up to unitary equivalence. In that case we speak of the
  {\em minimal Stinespring dilation} of $T$.

Consider now the set of all CP maps between $\cB(\cH_1)$ and
$\cB(\cH_2)$, for some Hilbert spaces $\cH_1$ and $\cH_2$. This set
can be partially ordered in the following way. Given two CP maps $T_1$
and $T_2$, we will say that $T_1$ is {\em completely dominated} by
$T_2$ \cite{bs} (and write $T_1 \le T_2$) if $T_2 - T_1$ is also a CP
map. If $T_1 \le \lambda T_2$ for some positive $\lambda \in \R$, we
will say that $T_1$ is {\em completely $\lambda$-dominated} by $T_2$.

Given a CP map $T$, the set of all CP maps completely dominated by $T$
can be characterized completely using a theorem of Belavkin and
Staszewski \cite{bs}, which is referred to as the ``Radon-Nikodym
theorem'' for CP maps, and asserts the following. Let $(V,\cE)$ be the
minimal Stinespring dilation of a CP map $T$. Then a CP map $S$ is
completely $\lambda$-dominated by $T$ if and only if it has the form
\begin{equation}
S(\rho) = V^*(\rho \tp F)V
\label{eq:rn}
\end{equation}
for some positive operator $F \in \cB(\cE)$ with $\norm{F} \le
\lambda$. Furthermore, the operator $F$ is determined uniquely by $S$
and $(V,\cE)$.

Next we would like to show that the ``Jamio\l kowski isomorphism''
between CP maps of $\cB(\cH_1)$ into $\cB(\cH_2)$ and the positive
operators on $\cH_2 \tp \cH_1$ is a direct consequence of the above
Radon-Nikodym theorem \cite{rag}.

We will consider quantum channels from $\cB(\cH_1)$ into $\cB(\cH_2)$,
where $\cH_1$ and $\cH_2$ are finite-dimensional Hilbert spaces. Let
us fix an invertible density operator $\rho$ on $\cH_1$, which we will
call the {\em reference state}. Let $p_i$ and $\phi_i$ be the
eigenvalues and the eigenvectors of $\rho$, and let us also fix an
orthonormal basis $\set{f_\mu}$ for $\cH_2$. Denoting by $\cE$ the
tensor product $\cH_2 \tp \cH_1$, define the isometry
$\map{V_\rho}{\cH_2}{\cH_1 \tp \cE}$ by
\begin{equation}
V_\rho \psi := \sum_{i,\mu}p^{1/2}_i \braket{f_\mu}{\psi}\phi_i \tp
f_\mu \tp \phi_i.
\label{eq:iso}
\end{equation}
Consider now a channel $T$. Let us define the unit vector $\Omega_\rho \in
\cH_1 \tp \cH_1$ by $\Omega_\rho := \sum_i p^{1/2}_i \phi_i \tp \phi_i$ and
the positive operator $F_{T,\rho}$ on $\cE$ by
\begin{equation}
F_{T,\rho} := (\idty \tp \rho^{-1})T \tp \id
(\ketbra{\Omega_\rho}{\Omega_\rho})(\idty \tp \rho^{-1}).
\label{eq:ft}
\end{equation}
With these definitions, we can write
\begin{equation}
T(\sigma) = V^*_\rho (\sigma \tp F_{T,\rho})V_\rho,
\label{eq:rnt}
\end{equation}
where the action of the coisometry $V^*_\rho$ on the
elementary tensors $\xi \tp \eta \in \cH_1 \tp \cE$ is given by
\begin{equation}
V^*_\rho (\xi \tp \eta) = \sum_{i,\mu}p^{1/2}_i \braket{f_\mu \tp
  \phi_i}{\eta} \braket{\phi_i}{\xi}f_\mu,
\label{eq:coiso}
\end{equation}
and then extended to all of $\cH_1 \tp \cE$ by linearity.

It is now an easy consequence of the Belavkin-Staszewski theorem \cite{bs} that the operator $F_{T,\rho}$ uniquely determines the
channel $T$, and that for any positive operator $F \in \cB(\cE)$, the map
\begin{equation}
M(\sigma) = V^*_\rho (\sigma \tp F)V_\rho
\label{eq:arb}
\end{equation}
is completely positive. We see therefore that any invertible density operator
$\rho$ on $\cH_1$ gives rise to an admissible pure state $\omega
\equiv \ketbra{\Omega_\rho}{\Omega_\rho}$, in the sense that the mapping
$\omega[T] := T\tp \id(\ketbra{\Omega_\rho}{\Omega_\rho})$ is invertible. That
is, the image of any density operator $w$ on $\cE$ under the inverse
map $\rho^\sharp$ is given by
\begin{equation}
\rho^\sharp(w) = V^*_\rho \left(\bullet \tp (\idty \tp
\rho^{-1})w(\idty \tp \rho^{-1})\right) V_\rho.
\label{eq:inv}
\end{equation}
It is important to realize that, in general, $\rho^\sharp(w)$ is {\em
  not} a quantum channel, unless $w$ satisfies the additional
  consistency condition
\begin{equation}
\tr_{\cH_2}[(\idty \tp \rho^{-1/2})w(\idty \tp \rho^{-1/2})] =
  \idty_{\cH_1}.
\label{eq:cons}
\end{equation}
This will hold automatically in the quantum system identification setting
(see Fig.~\ref{fig:subs}), provided that there is no additional noise
in the apparatus. We note also that the Jamio\l kowski isomorphism is
a special case of this formalism \cite{rag}, and is obtained if we pick as the
reference state the maximally chaotic density operator $(\dim
\cH_1)^{-1}\idty_{\cH_1}$.

Before we go on, let us remark that the set of pure states on $\cH_1
\tp \cH_1$ obtained by ``purification'' of invertible density
operators on $\cH_1$ does not exhaust all possibilities for admissible
states. In a recent paper, D'Ariano and Lo Presti \cite{dlp2} have
constructed a wide class of admissible states, which includes as a
subset the states discussed here.

\section{The performance of quantum system identification
  procedures}\label{sec:perform}

In this section we will quantify the performance of quantum system
identification procedures as a function of the admissible state used
as an input to the unknown channel.

Consider two density operators $w_1$ and $w_2$ on $\cH_2 \tp \cH_1$
that satisfy the consistency condition (\ref{eq:cons}). Then there
exist quantum channels $\map{T_1,T_2}{\cB(\cH_1)}{\cB(\cH_2)}$ such that
\begin{equation}
w_i = \rho[T_i] \equiv T_i \tp \id(\ketbra{\Omega_\rho}{\Omega_\rho}), \qquad i
= 1,2.
\label{eq:invch}
\end{equation}
Furthermore, from (\ref{eq:inv}) it follows that
\begin{equation}
T_i = \rho^\sharp(w_i) \equiv V^*_\rho(\bullet \tp F_i)V_\rho,
\label{eq:invch2}
\end{equation}
where $F_i := (\idty \tp \rho^{-1})w_i(\idty \tp \rho^{-1})$,
$i=1,2$. To determine how close the reconstructed channels
$\rho^\sharp(w_1)$ and $\rho^\sharp(w_2)$ will be when the
corresponding density operators $w_1$ and $w_2$ are close (say, in
trace norm), we will get a lower bound on the {\em channel fidelity}
\cite{rag2} between $\rho^\sharp(w_1)$ and $\rho^\sharp(w_2)$, defined
in the following way. Consider two channels
$\map{T_1,T_2}{\cB(\cH_1)}{\cB(\cH_2)}$, and define the density operators
\begin{equation}
\sigma_i := T_i \tp \id (\ketbra{\Omega}{\Omega}),
\label{eq:sigma}
\end{equation}
where $\Omega := (\dim \cH_1)^{-1/2}\sum_i e_i \tp e_i$, the summation
taken over some orthonormal basis of $\cH_1$. Note that
$\ketbra{\Omega}{\Omega}$ is an admissible state corresponding to the
maximally chaotic density operator $(\dim
\cH_1)^{-1}\idty_{\cH_1}$. Then the channel fidelity \cite{rag2} is
defined by
\begin{equation}
\cF(T_1,T_2) := \left( \tr \sqrt{\sigma^{1/2}_1 \sigma_2
\sigma^{1/2}_1}\right)^2,
\label{eq:chfid}
\end{equation}
where the quantity on the r.h.s. of (\ref{eq:chfid}) is the {\em
  mixed-state fidelity} \cite{joz}. We do not need all of the
  properties of the channel fidelity (\ref{eq:chfid}) [but see
  Ref.~\cite{rag2}], except the following:
\begin{equation}
2 - 2 \left(\cF(T_1,T_2)\right)^{1/2} \le \trnorm{\sigma_1 - \sigma_2},
\label{eq:fvdg}
\end{equation}
which is a simple corollary of the results of
  Fuchs and van de Graaf \cite{fvdg}. We also note that the channel
  fidelity has the natural property that $\cF(T_1,T_2) = 1$ if and
  only if $T_1 \equiv T_2$ (this is a straightforward consequence of the
  properties of the mixed-state fidelity \cite{joz}).

We can rewrite $w_1$ and $w_2$ from (\ref{eq:invch}) in terms of
$\rho$, $\sigma_1$, and $\sigma_2$:
\begin{equation}
\sigma_i = \frac{(\idty \tp \rho^{-1/2})w_i (\idty \tp
\rho^{-1/2})}{\dim \cH_1}, \qquad i=1,2.
\label{eq:sigma_rewrite}
\end{equation}
Then we can use the well-known inequalities $\trnorm{AB} \le
\norm{A}\trnorm{B}$ and $\norm{A^*A} = \norm{A}^2$, where
$\norm{\bullet}$ is the usual operator norm \cite{rs}, to get
\begin{equation}
\trnorm{\sigma_1 - \sigma_2} \le \frac{\norm{\rho^{-1}}\cdot \trnorm{w_1 -
    w_2}}{\dim \cH_1}.
\end{equation}
Combining this estimate with (\ref{eq:fvdg}), we obtain
\begin{eqnarray}
&& 2 -
    2\sqrt{\cF(\rho^\sharp(w_1),\rho^\sharp(w_2))}\nonumber
    \\
&& \qquad \qquad \le \frac{\norm{\rho^{-1}}\cdot \trnorm{w_1 -
    w_2}}{\dim \cH_1}.
\label{eq:step1}
\end{eqnarray}
Upon rearranging, we get the desired lower bound:
\begin{eqnarray}
&& \cF(\rho^\sharp(w_1),\rho^\sharp(w_2)) \nonumber \\
&& \qquad \ge \left
(1-\frac{\norm{\rho^{-1}}}{2 \dim \cH_1} \trnorm{w_1 - w_2}\right)^2.
\label{eq:lobound}
\end{eqnarray}
We see right away that worst-case performance of the channel
reconstruction procedure is controlled by the smallest eigenvalue of
$\rho$ (or, equivalently, by the largest eigenvalue of $\rho^{-1}$). This
fact has also been pointed out by D'Ariano and Lo Presti \cite{dlp},
and Eq.~(\ref{eq:lobound}) gives the corresponding quantitative
estimate. Note that in the case of $\rho = (\dim
\cH_1)^{-1}\idty_{\cH_1}$, the constant in front of the trace norm on
the r.h.s. of (\ref{eq:lobound}) is 1/2, which yields worst-case
performance that depends only on the states $w_1$ and $w_2$.

Note that we have discussed here the ideal scenario, namely that there
is no additional noise in the apparatus used for the channel
reconstruction. Any such disturbance will, of course, further degrade
the performance of the scheme.

\section{Discussion and conclusions}\label{sec:disc}

We have outlined a general mathematical framework for quantum system
identification, i.e., the determination (or estimation) of quantum
channels through their action on suitably chosen input density
operators (we have called them {\em admissible states}). In general,
the channel reconstruction procedure will involve the preparation of
an {\em entangled state}, followed by the application of an unknown
channel to one of the subsystems, leaving the other one intact. One
can show \cite{dlp,dlp2,dlpp} that the use of entangled states results in an
overall improvement, in either the precision or the stability of the
reconstruction procedure. On a more fundamental level, however, the
use of entanglement is also essential in light of the one-to-one
correspondence between quantum channels and bipartite density
operators [that satisfy the consistency condition (\ref{eq:cons})],
which can be explained in abstract terms within the framework of
Radon-Nikodym type theorems for CP maps \cite{bs,rag}.

In this paper we have emphasized quantum channels acting on
finite-dimensional algebras, in order to keep the presentation
simple. However, it is important (e.g., for quantum
information-theoretic studies in quantum optics) to have a
mathematical theory of quantum system identification in infinite
dimensions. Some steps in this direction have already been taken (see,
e.g., D'Ariano and Lo Presti \cite{dlp2} or Raginsky \cite{rag}). Let
us briefly comment on some of the big points. One starts with a
density operator $\rho$ that is invertible; however, the inverse is
now an {\em unbounded} operator. This implies that the reconstruction
map $\rho^\sharp$ will fail to be continuous, which will result in an
unbounded growth of statistical errors during the tomographic
estimation of the matrix elements of the Radon-Nikodym density $F_{T,\rho}$.

\subsection*{Acknowledgements}

This work has been supported by the Defense Advanced Research Projects
Agency and by the U.S. Army Research Office.

\end{document}